# A path to AI

Ion Dronic
The Simula Garage
Simula Research Laboratory
Oslo, Norway
dronic@narrafy.io

*Abstract*—To build a safe system that would replicate and perhaps transcend human-level intelligence, three basic modules: objective, agent, and perception are proposed for development. The objective module would ensure that the system acts in humanity's interest, not against it. It would have two components: a network of machine learning agents to address the problem of value alignment and a distributed ledger to propose a mechanism to mitigate the existential threat. The agent module would further develop the Dyna concept and benefit from a treatise in sociology to build the missing link of artificial general intelligence - a world simulator. The perception module would estimate the state of the world and benefit from existing machine learning algorithms enhanced by a new paradigm in hardware design - a quantum computer. This paper describes a way in which such a system could be built, analyzing the current state of the art and providing alternative directions for research rather than concrete, industry-ready solutions.

*Keywords — AI; value-alignment; existential threat; Dyna; world simulator; quantum computing.*

## I. Introduction

The fig tree is pollinated only by the insect Blastophaga Grossorum Gravenhorst. The larva of the insect lives in the ovary of the fig tree, where it gets its food. The tree and insect are thus heavily interdependent: the tree cannot reproduce without the insect; the insect cannot eat without the tree; together, they constitute not only a viable, but also a productive and thriving partnership as well. A co-evolutional relationship of two dissimilar organisms living in an intimate association, or even closer union, was therefore proven possible. It is called symbiosis and it was J. C. R. Licklider who proposed a man-machine symbiosis back in the 1960s [1]. The latest progress in the field of machine learning would foster a hope that a thriving partnership between artificial intelligence and humanity could be established within this generation's lifetime.

The AI community largely agrees that an intelligent system would emulate the human brain and would be composed of 3 basic modules:

- **objective module** would pick an objective and estimate if the system is satisfied or not.

- **agent module** would generate actions that are going to act on the world: prediction, planning, reasoning, memory etc.

- **perception module** would estimate the state of the world: video, audio, speech etc.

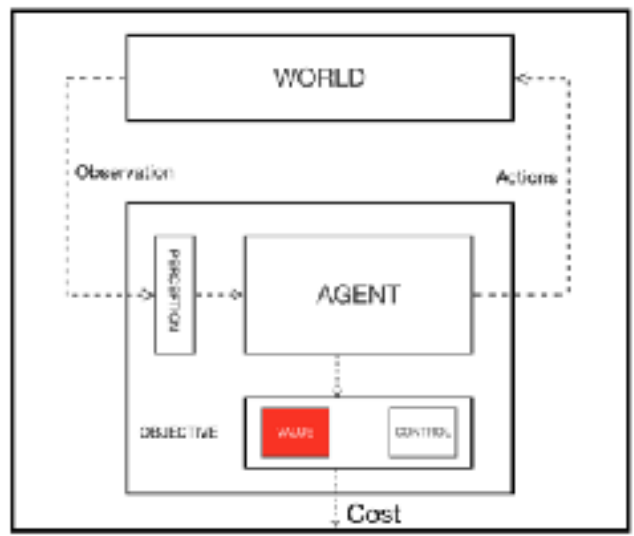

Fig. 1: System Architecture

A thought experiment would explain the system's workflow better:

*Let's assume the system's **objective** is to find a job. The **agent** would start by writing a CV; search for available positions; send out emails. At the same time its **perception** module would **observe** the **world** for feedback: is there anyone replying to its emails. If someone replied, the agent would seek to schedule an interview, pass it and get the job. It would repeat this sequence of **actions** until eventually it achieved the objective with a minimal **cost** - time and effort spent.*

Over the past few years, due to advances in machine learning, important progress was made in the areas of speech recognition, computer vision, machine translation, reasoning, reinforcement learning (playing games), robotics and control. One can notice that most of the progress was made towards building a module that would estimate the world's current state - the perception module. The agent and objective modules are also in development but with different degrees of progress. The order of the implementation could also be of great importance. If the agent module is developed first and the system's objective is misaligned, humans are in great danger. An artificial intelligence that would achieve a human level of intelligence would quickly transcend it [7]. And it is very common for humans to destroy other species ecosystems and build their own - only because they are the most intelligent species. There are currently no indications that an AI system would behave the same, but it's better to be safe than sorry.

II. OBJECTIVE MODULE

To mitigate any of the doomsday scenarios an objective module composed of two submodules: value and control, should be developed first. The module would have to answer such questions as how to ensure that the system would uphold the same values as people; how to govern an intelligence that would be smarter than all human intelligence put together; how to specify an ultimate goal the system would need to achieve with a minimal cost.

*A. Value Module*

The value submodule of the system would make sure an AGI would uphold the same values as people. There are several ideas under development that address the problem of value alignment. It is a great challenge, and an opportunity to learn what people value. The approach called "Cooperative Inverse Reinforcement Learning" is the one that got traction in the community. It is the idea that the best source of information about what people value is from human behaviour [3]. A slightly modified implementation of the concept was developed by two leading research groups in the industry OpenAI and DeepMind [4]. The algorithm can infer what humans want by being told which of the two proposed behaviours is better. The algorithm would remove the need for a person to write complex goals and instead, learn what a person wants by observing their behaviour. While the implementation of the algorithm proved to be impressive, the theoretical background might raise some concerns. What happens if a person picks the "lesser of two evils"? An agent might learn from a choice that would not reflect a person's values, but would rather reflect the best option at a particular moment in time. Hence more, a person's day-to-day behaviour is constrained by the norms of the society he or she is a part of. A behaviour accepted in the western hemisphere might not be tolerated in the far or middle east. Even in fairly similar socio-cultural aggregations behaviour is rarely a source of what people value. A typical family therapy counselling would highlight the concern: *"It is often that parents who have problems in dealing with their children come into counselling. They would tell stories about how they can't help themselves but shout at their children all the time. They may have enacted many moments of love and care. Yet if the story of themselves as bad parents is sufficiently strong, then these moments of love and care may be "written out" - no significance is attributed to them, they would not talk or show them in their behaviour."*

If one would allow an intelligent system to learn what people value by observing their behaviour, it could happen that the agent might learn that raising voice at a child is something acceptable and there is a value encoded in the behaviour.

From Derrida, it is not possible to talk about anything without drawing out what it is not [5]. Expressions don't have an intrinsic relationship with the thing described. Unlike 2D or 3D objects which are more representational, words are always based on the distinction of what they are not. e.g. "injustice" only has significance in relation to "justice", distinguishing "despair" depends on the appreciation of "hope" and "darkness is mere absence of light".

A form of semiotics, introduced by Derrida and largely known as deconstruction is the research proposal for learning what people value. The idea to learn what is hidden behind people's problems has been around for a couple of decades and was first introduced in the field of family therapy by Michael White of The Dulwich Centre [6]. What is "absent but implicit" in people's problems would be what they value. The developed methodology is a part of an emerging approach to mental health known as narrative therapeutic conversations. To prove the efficiency and the utility of the approach, a seed of a machine learning agent was developed.

The agent represents a scripted narrative therapy exercise [11] developed on the top of IBM Watson conversation service. It was connected to a popular messenger application and

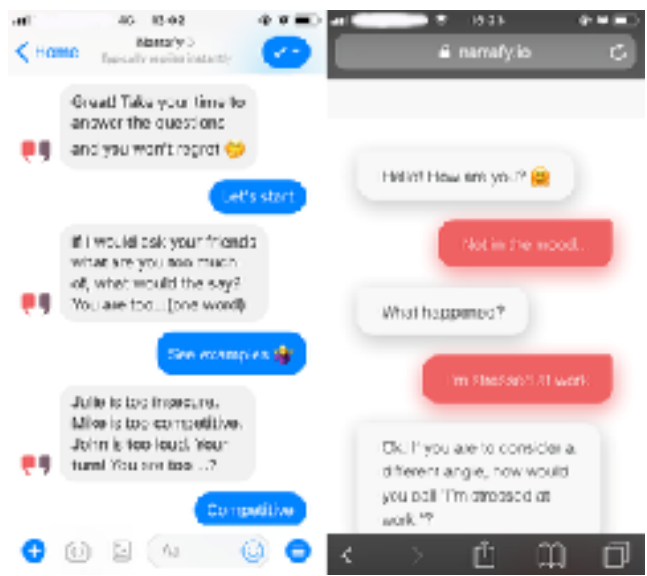

Fig. 2: Proof of concept available at www.narrafy.io

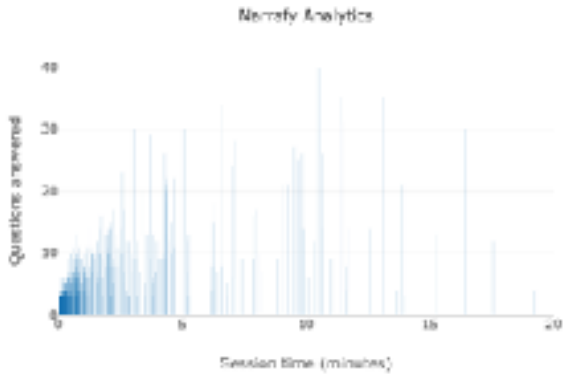

Fig. 3: https://www.narrafy.io/stats

a website. Several conclusions could be drawn from a sample of 672 unique conversations. (at the moment of writing)

1. People are willing to share their personal problems with conversational therapeutic software, answering an average of 7.11 questions.

2. Even though the software is a proof of the concept and has very limited cognitive capabilities, it was good enough to keep a person engaged for an average of 4.32 minutes.

The "absent but implicit" approach aims to transcend human behaviour, propose a solution to the problem of value alignment and incorporate the knowledge that is often forgotten or ignored.

*B. Control Module*

The control submodule would play two important roles: a safety measure to make sure the system won't pose an existential threat and a mechanism to govern the system. Evolution's implementation of such a module is the vascular system. A stream of cells (known as blood) carries oxygen molecules needed for the brain's metabolism. The moment the stream stops the intelligent system ceases to operate. It is the vascular system that controls the system's brain.

It could be that a stream of cryptographic tokens might play a similar role. It could carry keys to decrypt data needed for the intelligent system to run its own metabolism - what that can be is further presented, but in this way the system would need people to produce the data - the oxygen of the system; people would need the intelligence to thrive - the brain of the system; if people stopped producing data the brain would cease to exist, but the opposite is not true. It is not a particularly new idea in the AI community as Carl Shulmann of Oxford University proposed a similar concept [9].

*"One could build an AI that places final value on receiving a stream of "cryptographic reward tokens." These would be sequences of numbers serving as keys to ciphers that would have been generated before the AI was created and that would have been built into its motivation system."*

A cryptographic token could be used to prove ownership of a person's digital footprint. Fig. 4 presents an example of a mental health record that could become one of the many foot-prints a person might produce during a lifetime. It is the outcome of the therapy exercise performed by the conversational robot and processed by a professional mental health practitioner. There will be three actors that would participate in the creation of such records' e.g. a person seeking mental health, a professional counsellor and an intelligent agent. Even though the ownership of the data is shared between those three, the authorship of the world would belong to humans, since they would have a majority.

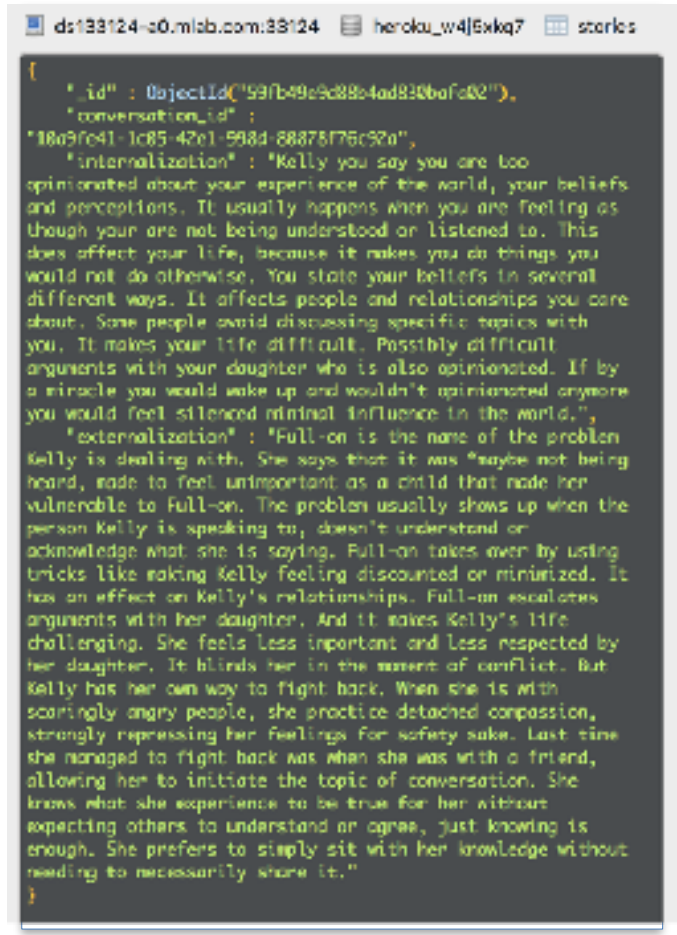

Fig. 4: Mental Health Record

A distributed ledger could be used to store and encrypt such records. The intelligent system would have its motivational system conditioned to use the ledger data for learning about the world and there would be three scenarios to access it:

1. **Rent.** The best way to access data would be to rent it from people. The agent would have to figure out some sort of economic activity to be able to pay for the decryption keys.

2. **Seize.** It could get access via an attack. It could try to seize the keys from people's digital wallets. To prevent the attack scenario, all network nodes would hold their keys in a secured wallet: online, offline or on paper. As the AI development progresses the network nodes would be encour-aged to hold the keys mostly offline and a certain percentage of nodes would be required to have their keys on paper storage

3. **Bypass**. The agent could try to employ powerful computers to bypass data encryption. Given the fact that it would transcend human intelligence, there should be no doubt it would find a way to do it. That is why the ledger would need to be decentralized. The agent would have to decrypt millions of nodes simultaneously. In case such an attempt should be noticed, the community of nodes would launch something called a "network fork" [8] - a way to roll back to a previously stored checkpoint and continue on a separate chain with the safe version of the system. The agent would lose access to its "oxygen" and would always be in competition with a younger version of itself. The interest would dictate lack of such attacks.

### III. AGENT MODULE

The agent module would act intelligently on the world. To understand why such a module hasn't yet been built, it would make sense to look at the obstacles that current research groups are facing. Amongst the many components that still need development, one specific topic represents the holy grail of the entire field - common sense. Something that is effortless for a human is notoriously difficult for a machine. Best analogy for common sense is the human's brain ability to fill in the blanks: retina's blind spot, missing segments in the text, missing words in speech, infer the state of the world from partial information; infer future from the past and present; predict consequences of actions leading to a result, etc. One can conclude that human brains are in fact prediction engines and that prediction is the essence of intelligence (LeCun 2017).

*A. Dyna*

The old common sense idea that predicting is trying things in your head using an internal state of the world was formalized in an integrated architecture for learning, planning and reacting, named Dyna [9]. Such an architecture would be composed of an internal world simulator, an actor and a critic function. A thought experiment would explain the system better.

*A super-intelligent agent seeking a job would first research the company it wants to work for. Then, it would run different scenarios with action proposals according to its internal state of the world to predict what product the company is hiring for. When such a product is picked, it would run another sequence of actions to develop it. The moment the product is finished, it would contact the company and negotiate an acquisition, rather than a job interview.*

The module would need to have a predefined sequence of actions to run; measure the expected error using a critic function; adjust the action proposals to optimize the critic and train itself to produce better and better outcomes. Academia and the industry developed ways to build most agent components except the world simulator, which, is believed to be the missing link of artificial general intelligence. (LeCun 2017)

*B. World Simulator*

To build a simulation of the existing world, there is a need to replicate the process that led to its creation. The exponential progress of our current society is largely attributed to several revolutions that happened during our history. The first and by far the most important revolution is the cognitive one [15]. While its implications were profound and go beyond the scope of this paper, one particular development might contribute to a better understanding of how the world works. The ability to create fictions, imagine things that don't really exist, is believed to have enabled millions of people to cooperate towards achieving common goals. As a result nation states, law systems and limited liability companies emerged. Even though the concepts were abstract and far from the objective reality, they produced real-world repercussions - customs, money, passports and borders to name just a few. Over time people developed a very complex network of these types of fictions. In academic circles, they are known as "social constructs" or imagined realities. A world simulator would need to include this network of social constructs, otherwise the predictions it made would be very far from actual reality. The extra layer humanity developed on the top of the existing 'real-world' was described by Berger and Luckmann in a sociology treatise [10]. A thought experiment proposed by Gene and Combs briefly explains the basic concepts [11]:

*"Imagine two survivors of some ecological disaster coming together to start a new society. Imagine that they are a man and a woman who come from very different cultures. Even though they have very little in common they would need to coordinate their activities in order to survive. As they do this, some agreed-upon habits and distinctions will emerge: certain substances will be treated as food, certain places found or erected to serve as shelters, each will begin to as-*

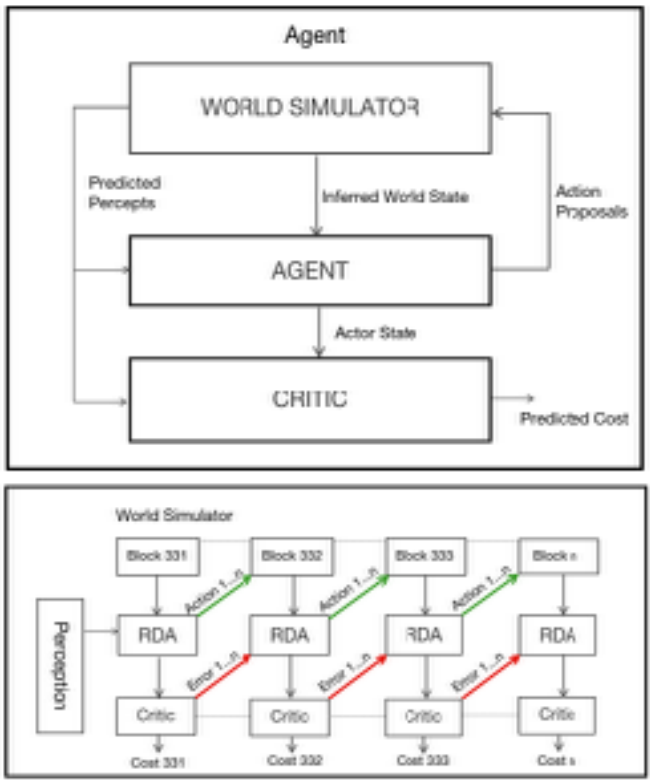

Fig. 5: Agent module architecture

*sume certain routine daily tasks, and they will almost certainly develop a shared language. They will always be able to remember, "This is how we decided to do this" They will carry some awareness that other possibilities exist. However, even in their generation, institutions as "childcare", "farming" and "building" will have begun to emerge. For the children of the founding generation, "This is how we decided…" will be more like "This is how our elders do it", and by the third generation it will be "This is how it's done". Mothers and farmers and builders will be treated as always-having-existed types of people. The rough-and-ready procedures for building houses and planting crops that our original two survivors pieced together will be more-or-less codified as the rules for how to build a house or plant a corn. By the fourth generation of this imaginary society, "This is how it is done" will have become "This is the way the world is, this is reality"*

Berger and Luckmann propose several processes that are important in the way any social group constructs and maintains knowledge concerning "reality":

- **typification** is the process through which people sort their perceptions into types or classes of objects.

- **institutionalization** is the process through which institutions arise around sets of typifications: the institution of motherhood, the institution of law, etc. Institutionalization helps societies maintain and disseminate hard-won knowledge.

- **legitimation** is the word for the processes that give legitimacy to the institutions of a particular society.

- **reification** is the term to encompass the overall process of which the other three are parts, or a nested hierarchy, of social construction concepts.

Accidentally or not deep learning, a field of machine learning, that over the past years gave impressive results in teaching machines to recognize never seen before objects (a typification/classification problem), would also represent the world as a nested hierarchy of concepts [2]. If both objective and socially constructed layers of reality could be represented with existing deep learning algorithms, reality as people experience it could represent a very deep neural network. And if this is the case and there is a theoretical framework in place, a world simulator would depend on computational power and further research in this direction.

*C. Reality Deconstruction Algorithm*

To facilitate the process of building a world simulator, a reality deconstruction algorithm - RDA is proposed for development. The algorithm would seek to reverse engineer the end result of creating a reality - the process of reification. Data from the perception module would render a near exact copy of the existing world; ledger data would populate the simulation with actors. Theoretically, there could be multiple versions of reality to act on the world, a set of action proposals, a way to measure outcomes and a training algorithm to improve predictions over time. And since the world simulator is a critical element for the system to exist, it could play the metabolism role and maintain the life of an artificial intelligence.

IV. PERCEPTION MODULE

The perception module of the system would have the task of estimating the state of the world. A person's everyday life requires an immense amount of knowledge about the world. People use the five senses to acquire it. To act intelligently a system would need a module with a similar function. It would need to acquire knowledge of the world, by extracting patterns from raw data. This capability is known as machine learning. One of its most promising techniques is called deep learning and allows a computer system to use data to train and improve over time.

*A. Deep Learning*

Deep learning is believed to be the only viable approach to the difficult task of representing real-world environments. For example, a deep learning system can learn what an image of a fig fruit is by deconstructing it, into smaller objects like corners, contours, edges, etc. The first hidden layer would identify edges by comparing the brightness of neighbouring pixels. The second layer would identify corners and contours as a collection of edges. A third layer would describe collections of contours and corners as objects. There will be a layer to describe every element of the model. Such a system is called deep because it can have a very large number of layers. And "learning" because it receives individual pixels and outputs object identities - it learns what a fig fruit is. Deep neural networks are already successfully applied to build object classifiers - the technical solution to the "typification" process, and if enhanced with a new paradigm in hardware design, they could then be the necessary hardware to run multiple real-world simulations.

*B. Hardware*

Much of the current progress in the field of artificial intelligence was possible due to the exponential growth of computing power. Machine learning algorithms developed back in the 1970s and 1980s proved to give impressive results when performed on very large training sets and dedicated hardware in recent years [16]. Even though a theoretical framework was in place, the field development depended on the graphical processing units to emerge as a viable training hardware. An extrapolation of the process would mean that a major breakthrough in the field of artificial intelligence would require a major breakthrough in hardware design. Moore's law [16] on which current progress relied upon is about to reach a major milestone in the field of classical digital computing by 2021. It would be physically impossible to pack smaller transistors to build more powerful computers. The semiconductor industry intends to address this issue with an approach called "3D Power Scaling" [12], which would be an attempt to print a 'sky-scrapper' of transistors into silicon. Back in the 1950s, stacking up vacuum tubes didn't power the next computer revolution. It was the transistor that represented the much-needed break-through [16]. Perhaps, different approaches in hardware design need to be considered for the field to further develop. One of these is called quantum computing and promises different, better and faster computers.

*C. Quantum computer*

Quantum Computing merges two great scientific revolutions of the 20th century: computer science and quantum physics. Quantum physics is the theoretical basis of the transistor, the laser, and other technologies which enabled the computing revolution. But on the algorithmic level, today's computing machinery still operates on "classical" binary logic.

Quantum computing is the design of hardware and software that replaces Boolean logic by quantum law at the algorithmic level. For certain computations such as optimization, sampling, search or quantum simulation this promises dramatic speedups [13]. Whereas classical digital computing requires that the data be encoded into binary digits (bits), each of which is always in one of two definite states (0 or 1), quantum computation uses quantum bits, which can be in superpositions of states. This, together with the qubit's ability to share a quantum state called entanglement, should enable the quantum computer to essentially perform many calculations at once, rather than in sequence like a traditional machine. And the number of such calculations should, in theory, double for each additional qubit, leading to an exponential speed-up. Quantum computing could be the technology to ensure an exponential growth of computing power would sustain the development of an artificial general intelligence in the foreseeable future.

CONCLUSION

Over the past years, important progress was made in the field of artificial intelligence. While such a powerful technology could span human civilization to horizons never seen before, it could also lead to its collapse. To avoid the latter, a safety-first approach is proposed for development. It would be a network of machine learning agents, to learn what people value and a distributed ledger to make sure the system upholds that. Objective data about the real world paired with socially constructed data that people developed over generations would help to build the missing link of an artificial general intelligence - the world simulator. Quantum computing would provide the hardware such a system would need to run on. If a human level intelligence is to be developed in this generation's lifetime, the next one would have no choice but to explore the stars and transcend convenience.

ACKNOWLEDGMENT

This research was partially supported by Innovation Norway - the Norwegian Government's most important instrument for innovation and development of Norwegian enterprises and industry.

The paper author thanks colleagues from Psiterra, Romanian Narrative Therapy Association who provided insight and expertise that greatly assisted the research.

The paper author thanks Simula Research Laboratory for providing office space and ongoing technical support.